Title: Piezoresistance in Silicon and its nanostructures

Author: A. C. H. Rowe

Affiliation: Physique de la matière condensée, Ecole Polytechnique, CNRS

Address: Route de Saclay, 91128 Palaiseau, France

Email: alistair.rowe@polytechnique.edu

Telephone: +33 (0)1 6933 4787

Best figure representation of work: Figure 4

Abstract:

Piezoresistance (PZR) is the change in the electrical resistivity of a solid induced by an applied mechanical stress. Its origin in bulk, crystalline materials like Silicon, is principally a change in the electronic structure which leads to a modification of the charge carriers' effective mass. The last few years have seen a rising interest in the PZR properties of semiconductor nanostructures, motivated in part by claims of a giant PZR in Silicon nanowires more than two orders of magnitude bigger than the known bulk effect. This review aims to present the controversy surrounding claims and counter-claims of giant PZR in Silicon nanostructures by summarizing the major works carried out over the last 10 years. The main conclusions to be drawn from the literature are that i) reproducible evidence for a giant PZR in un-gated nanowires is limited, ii) in gated nanowires giant PZR has been reproduced by several authors, iii) the giant effect is fundamentally different from either the bulk Silicon PZR or that due to quantum confinement, the evidence pointing to an electrostatic origin, iv) released nanowires tend to have slightly larger PZR than un-released nanowires, and v) insufficient work has been performed on bottom-up grown nanowires to be able to rule out a fundamental difference in their properties when compared with top-down nanowires. On the basis of this, future possible research directions are suggested.

Article: following page.

## Section 1: Introduction and history of piezoresistance

To properly understand how and why the electrical resistance of solids changes with applied mechanical stress, and to fully understand just how different the apparent effects in nanostructures are compared to the bulk case, it is useful to begin by considering the simple undergraduate formula for the resistance:

$$R = \rho \frac{l}{A} \tag{1}$$

where $\rho$ is the resistivity, *l* is the length, and *A* is the surface area of a parallelepiped conductor. If this conductor is subject, for example, to a uniaxial tensile stress parallel to its length axis, then the length will increase to *l* + Δ*l* while the surface area will be reduced due to Poisson's effect from $A = w^2$ to $(w - \nu \Delta l)^2$ where $\nu$ is Poisson's ratio. In the limit of small strains, $\varepsilon = \Delta l / l << 1$, the resulting relative resistance change due entirely to the modified sample dimensions is approximately $(1 + 2\nu)\Delta l / l$. This yields a measure of the piezoresistance (PZR) called the gage factor, the relative resistance change per unit strain, given by:

$$G = \frac{\Delta R}{R} \frac{1}{\varepsilon} \approx 1 + 2\nu \, . \tag{2}$$

For most solids 0.3 < $\nu$ < 0.5 so that the geometric PZR is typically expected to lie in the range 1.6 < *G* < 2. In reality, this range corresponds to the minimum possible gage factor since $\rho$ itself also changes with applied mechanical stress. Strictly speaking, the PZR is defined as the change in $\rho$ with applied stress. A more physical definition of the PZR which does not depend on the sample dimensions is the $\pi$-coefficient, defined with respect to the applied stress (*X*) as:

$$\pi = \frac{\Delta \rho}{\rho} \frac{1}{X} \, . \tag{3}$$

When the geometric changes to a conductor due to applied stress are negligible, the $\pi$-coefficient of Eq. (3) is related to $G$ in Eq. (2) via Young's modulus, $Y$, since (at least in scalar form) $X = Y\varepsilon$.

At the beginning of the 20th century, the observation of a resistance change over and above the geometric change described above was difficult to explain. The fact that $\rho$ varies with applied stress is not predicted from Drude's classical model of electron motion. In this case the momentum relaxation time associated with the carrier mobility, $\tau_m$, varies due to applied pressure as the inverse of the charge carrier density – in the case of electrons, denoted $n$. As such the resistivity,

$$\rho = \frac{1}{n\mu_n q} = \frac{m^*}{nq^2\tau_m}, \tag{4}$$

no longer changes with stress (or in the case of the earliest measurement, hydrostatic pressure). Here $q$ is the absolute value of the electronic charge. Although the first measurements date back well into the 19th century, Bridgman was the first to systematically study PZR in a large selection of metals subject to hydrostatic pressure[1]. He found PZR coefficients ranging from slightly less than 1 x $10^{-11}$ $Pa^{-1}$ (in some of the transition metals), up to about 70 x $10^{-11}$ $Pa^{-1}$ (in the alkali metals). Here the modern sign convention is followed: compressive stresses are negative while tensile stresses are positive. A positive PZR coefficient under hydrostatic pressure (i.e. a negative stress), implies that for the majority of elements shown in white circles in Fig. 1(a), the resistance drops with increasing pressure. Figure 1(a) summarizes Bridgman's complete works on the measured pressure coefficients of resistance (for pressures less than 1 GPa) across the periodic table[2]. For each element in this figure, the radius of the circle around each symbol is proportional to the decimal logarithm of the measured PZR coefficient (i.e. $\log_{10} \pi$) so that, for example, the piezo-response of Sodium under hydrostatic pressure is a factor of 10 larger than that of Paladium. The clear-cut tendencies across the rows (and within the columns) of the periodic table demonstrates that the *piezoresistance is an electronic structure phenomenon*. Interestingly some metals such as Bismuth, Antimony, Lithium and Cesium, compounds of which are currently of interest for their unusual electronic properties[3], exhibit anomalous behavior where the resistance *increases* with pressure (shaded, gray circles). A correct explanation of PZR phenomena therefore requires a quantum mechanical description[i].

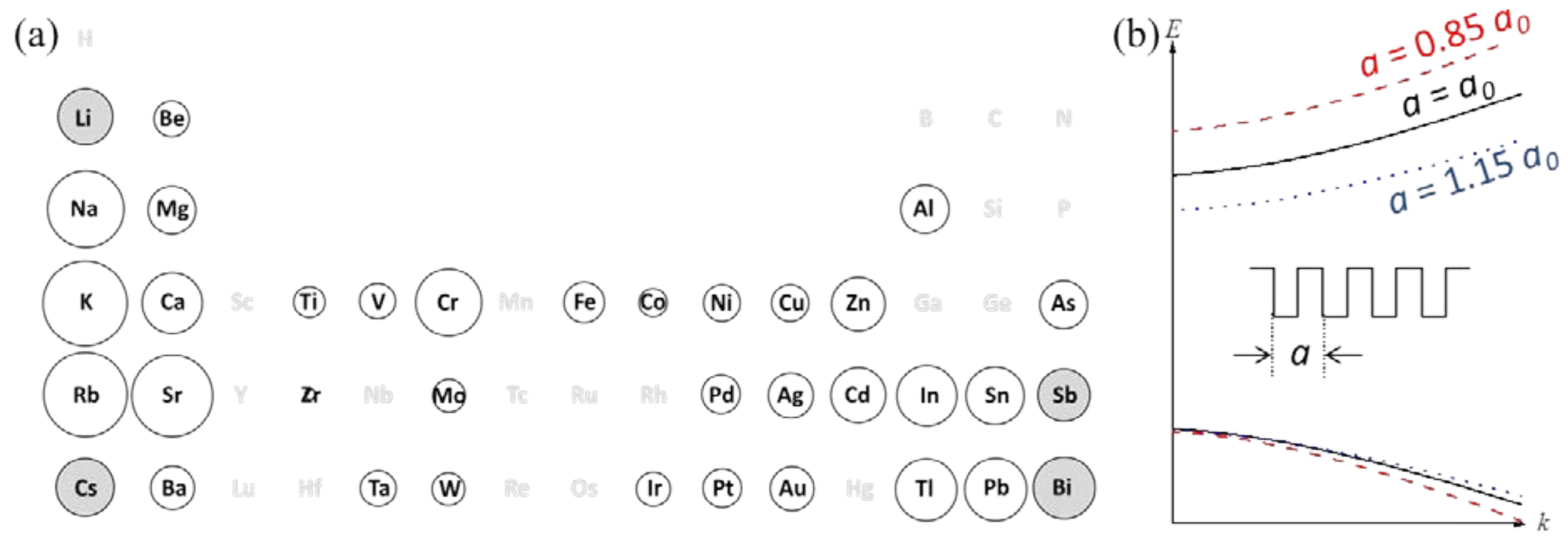

**Figure 1** (a) A graphical representation of the measured PZR coefficients (here under hydrostatic pressure) of a large number of metallic elements (from Ref. 1 and Ref. 2). The radius of the circle around each element is proportional to the logarithm of the measured coefficient. Un-shaded circles correspond to the usual negative coefficients (i.e. a reduction in resistance with increasing pressure) while the gray shaded circles correspond to anomalous increases in resistance with increasing pressure. Grayed-out elements correspond to materials that were not measured. The trends across the rows (and in the columns) of the periodic table demonstrate that PZR is an electronic structure phenomenon requiring a quantum mechanical description. (b) The effect of applied mechanical stress on electronic structure demonstrated by plotting the dispersion relations obtained from the one-dimensional Kronig-Penney model. With respect to the nominal zero stress case (solid, black line), 15 % tensile strained (dotted, blue line) and 15 % compressive strained (dashed, red line) cases exhibit a stress-induced change in the bandgap, the effective masses and the density-of-states. According to Eq. (6), each of these changes can in principle contribute to a stress dependence of the resistivity, $\rho$.

[i] Bridgman motivates his work by suggesting that the limitations of Drude's classical description of electron motion, and the need for a different (quantum) theory, would have been revealed earlier if more attention had been paid to measurements of resistance under pressure dating from the second half of the 19th century. Instead, a definitive demonstration of this had to wait until the more difficult measurement of the specific heat of solids at low temperatures in the early 20th century.

Armed with the tools of quantum mechanics it is immediately clear from even the simplest one-dimensional models of the crystal potential that mechanical stress should modify $\rho$. Consider the Kronig-Penney model in which a non-interacting electron moves in a periodic one-dimensional array of square potential wells spaced by the lattice parameter, $a = a_0$. The potential wells represent the atoms of the 1-dimensional lattice. As is well known, an analytical solution to Schrodinger's equation exists in this case and the dispersion relationship between energy, $E$, and wavevector $k$, is given by[4]:

$$\cos(ka) = \cos(\alpha a) + \frac{m}{\hbar^2 \alpha} \sin(\alpha a), \tag{5}$$

where $\alpha^2 = 2mE/\hbar^2$ and $m$ is the free electron mass. A bandgap in the electronic structure described by Eq. (5) is shown in Fig. 1(b) (black, solid curve). If $a$ is now reduced from its initial, unstressed value ($a_0$) to $0.85a_0$ to simulate a compressive stress resulting in a 15 % strain, the dispersion relations are modified according to the dashed, red lines in Fig. 1(b). This figure also shows (dotted blue lines) the effect of an increase of $a$ to $1.15a_0$ of its initial value which simulates an applied tensile stress resulting in 15 % strain. The dependence of Eq. (5) on $a$ results in a bandgap dependence on stress, an effective mass dependence on stress, and also a dependence of the density of states, $2\pi g = dk/dE$, on stress. It is then clear that the mobility $\mu_n = \mu_n(X)$ depends on applied stress due to the change in both the effective mass and in the momentum relaxation time, $\tau_m \propto 1/g$, with stress. Moreover the carrier concentration also becomes stress-dependent, $n = n(X)$, through its dependence on $g$ as well as on the bandgap. Thus, even ignoring changes to the electron dynamics induced by stress, it is clear that:

$$\rho = \rho(X) = \frac{1}{n(X)\mu_n(X)q} . \tag{6}$$

**Section 2: Piezoresistance in bulk Silicon**

Perhaps the most extreme, non-Drude-like case considered by Bridgman was that of the crystalline semiconductors, Germanium and Silicon[2]. Even a qualitative description of the observed behavior was difficult since “the resistance is susceptible to dislocations” and the “non-hydrostatic nature of the pressure” i.e. the crystalline nature of the material. While he was unable to draw any concrete conclusions regarding the properties of these materials, he did note that, in the case of Silicon, the sign of the $\pi$-coefficient changed depending on whether the crystal was doped n-type or p-type. In fact these observations came only two years before the classic experimental work of Smith that revealed the PZR effect in Germanium and Silicon in qualitative detail[5]. The results obtained by Smith were particularly striking in that, for Silicon under the appropriate conditions, the measured $\pi$-coefficients were more than twice as large as those measured in the alkaline metals Potassium and Rubidium (see Fig. 1(a)), although it was found to decrease with increases in the doping density sufficiently large to push the semiconductor into the degenerate regime[6].

Smith suspended small weights from mono-crystalline boules of Silicon and Germanium. By doing this for a variety of crystal orientations and for current flow either parallel or perpendicular to the applied, uniaxial tensile stress, he was able to estimate the $\pi$-coefficients for all configurations and to show that, in general, it is a tensor quantity[5]. He also showed, like Bridgman, that the dopant type significantly modifies the PZR. Although he offered only a phenomenological description of the effect, the microscopic description of the phenomenon in n-type Silicon where the electronic structure is relatively simple, was rapidly accomplished[7]. To understand this co-called charge-transfer model, it is useful to consider the case of a uniaxial compressive stress applied along the <100> direction in bulk, n-type

Silicon. The longitudinal PZR (i.e. when an electrical current is flowed parallel to the direction of the applied stress) will be considered.

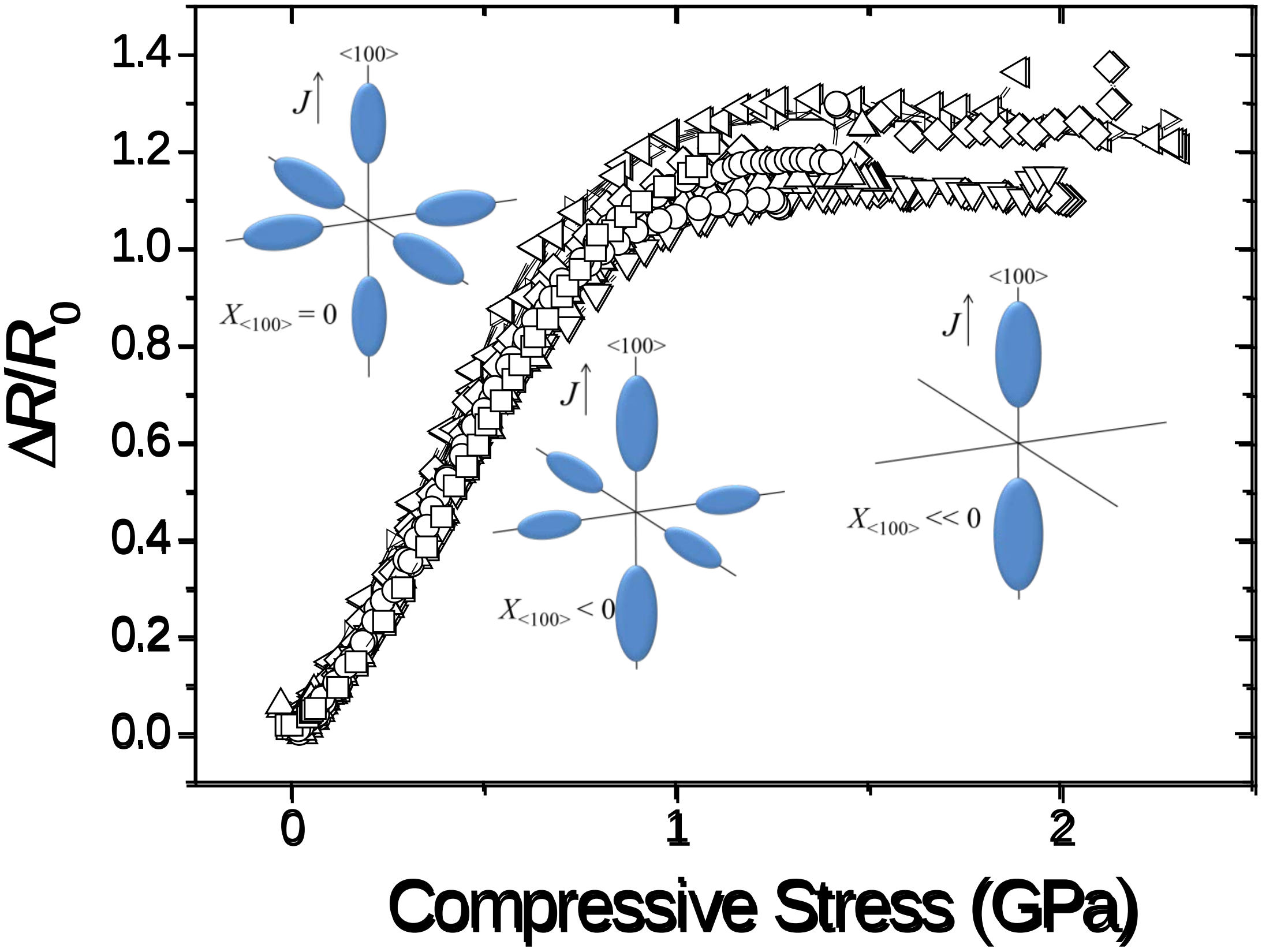

**Figure 2** Longitudinal PZR in n-type Silicon along the <100> crystal direction (adapted from Ref. 9). At zero applied stress the conduction band minimum consists of six equivalent valleys along the <100> directions that are degenerate and therefore equally populated. As the uniaxial compressive stress is increased, the degeneracy of the six valleys is lifted. The longitudinal valleys shift downwards in energy relative to the four transverse valleys resulting in a charge transfer to the two longitudinal valleys. This changes the density-of-states weighted effective mass since the parallel and perpendicular masses are not equal. Since the parallel mass is greater than the transverse mass this increases the resistance until all the charge is transferred from the transverse valleys. The diagrams (inset) schematize the charge transfer process.

Under zero applied stress, the conduction band minima of Silicon are approximately 83 % of the distance to the Brillouin zone boundary along the $\Gamma$X or {100} crystal directions. Consequently the iso-energetic surfaces are no longer spheres but degenerate ellipsoids of which there are six. These ellipsoids are characterized by two effective masses; the parallel and transverse effective masses which correspond to the major and minor axes of the ellipsoids respectively. When a compressive uniaxial stress is applied along the <100> direction the ellipsoids with the long axis oriented along the <100> direction (i.e. the longitudinal ellipsoids) are shifted down in energy with respect to the ellipsoids whose long axis is oriented parallel to the <010> and <001> crystal directions (i.e. the transverse ellipsoids). Electrons from the perpendicular valleys are then transferred to the parallel valleys so that the density-of-states effective mass changes. Since the parallel mass is larger than the transverse mass, according to Eq. (4) a measurement of the resistance along the <100> direction reveals an increase in its value with increasing applied compressive stress. This increase continues until all the electrons are transferred from the transverse to the longitudinal valleys[8] as indicated in Fig. 2, as has recently been verified up to unprecedentedly high stresses of the order of 3 GPa[9]. With respect to Eq. (6) therefore, the PZR in n-type Silicon is primarily due to a change in the $m^*m^*$ and thus the mobility, $\mu_n$, of conduction band electrons. Based on this description, the PZR is quite clearly dependent on crystal direction and a qualitative description of the effect in all directions can be obtained using Smith's treatment[5] and the method described by Kanda[10].

The valence band structure is more complex, with three bands in close (energetic) proximity near the band maximum at the center of the Brillouin zone. It was recognized early on that the charge transfer mechanism would therefore not describe the PZR in p-type Silicon[11]. In the intervening period between Smith's first measurements and the beginning of the 21$^{st}$ century a number of authors have treated and tried to identify the keys elements that describe PZR in p-type Silicon[12,13,14,15]. In addition to charge transfer between for example, heavy and light hole bands, their proximity also results in a band warping (i.e. a change in the curvature of each band as stress is applied) and potentially to changes in $\tau_m$ due to stress-induced modifications in the scattering rates. Surprisingly for a material whose electronic properties have been studied in detail since the 1950s, the preponderance of each of these effects is still to some extent debated, leading for example to conflicting predictions about the PZR at high stress[16,17]. In the simplest band-edge models, the PZR was predicted to saturate at high stress much like the n-type case[16] whereas solutions of the Boltzmann equation using the full band structure predict a much weaker, or even an absence of saturation, at least up to 4 GPa[17]. Recent experimental results along with *ab initio* band structure calculations demonstrate that the latter interpretation is probably correct[9]. The details of the longitudinal PZR along the <110> and the <111> crystal directions are of interest, both to engineers exploiting strained Silicon technology to increase the hole mobility[16], but also because these are the most common orientations of Silicon nanowires where the largest PZR is claimed[18].

Under a compressive uniaxial stress applied parallel to the <110> crystal direction, the light hole band moves to higher energies while the heavy hole band moves down. This is shown schematically inset in Fig. 3. In doing so a band anti-crossing appears, initially near the Brillouin zone center and moving outwards as the stress increases. This anti-crossing is extremely important for the PZR since the effective mass of holes in states near the anti-crossing is significantly different from either that of the heavy or the light holes. As the stress increases the anti-crossing moves out to higher wavevectors and into the tail of the hole distribution[9]. Below stresses of approximately 3 GPa there is still a non-negligible population of holes near this band anti-crossing and this results in very little drop off in the resistance change with applied stress in this range (see Fig. 3). The case of an applied stress along the <111> crystal direction is qualitatively similar to the <110> case. In general therefore, the PZR in p-type Silicon, like the n-type case, is due to a change in the transport effective mass.

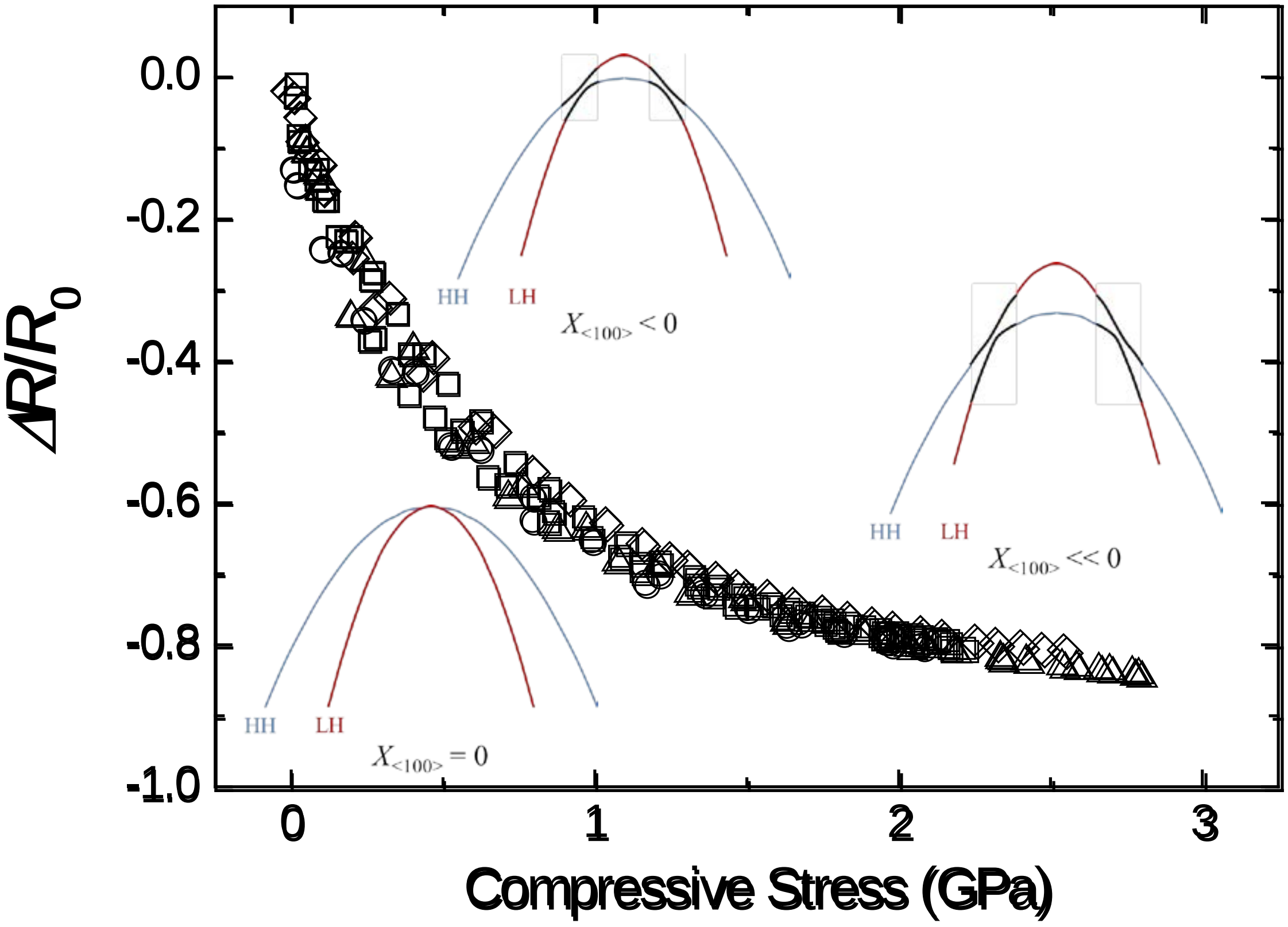

**Figure 3** Longitudinal PZR in p-type Silicon along the <110> crystal direction (adapted from Ref. 9). At zero applied stress the HH and LH bands are degenerate at the Brillouin zone center. As the uniaxial compressive stress is increased this degeneracy is lifted, with the LH (red) band moving upwards in energy relative to the HH (blue) band. The resulting anti-crossing (black) yields a significant fraction of holes whose transport effective mass is very different from either the HH or the LH mass. The population of holes near this anti-crossing, which moves to lower energies further from the center of the Brillouin zone with increasing stress, is still non negligible at room

temperature and at stresses near 3 GPa. Saturation of the PZR is therefore weak or non-existent. At higher stresses all the holes will reside in the top band and have an effective mass given by the LH mass. In the absence of any stress-induced changes in $\tau_m$, the PZR is then expected to saturate.

A largely overlooked point, which may be important for the understanding of the PZR of nanowires, is the non-linearity in the PZR response for both n-type and p-type Silicon. While the form and values of higher order PZR coefficients have been described and measured, the aim has usually been to address non-linearities at higher stress levels (i.e. in the 100 MPa range and above)[19,20]. In fact it is clear from the above discussion that the first order PZR coefficients should not be the same in compression and in tension. In the n-type case treated above, compressive stress corresponds to the transfer of electrons from the four transverse valleys *to* the two longitudinal valleys. Naturally, tensile stress will results in a transfer of electrons from the two longitudinal valleys to the four transverse valleys. The resulting variations in resistance with therefore be different in compression and in tension. Similarly, for p-type Silicon in the case treated above, compressive uniaxial stress shifts the light hole band upwards resulting in the band anti-crossing. Eventually all holes will reside in the warped light hole band. In tension the reverse occurs and the heavy hole band moves upwards in energy. There is no band anti-crossing and eventually all holes reside in the heavy hole band. Like n-type material, the stress response is asymmetric. As a result the $\pi$-coefficients can be significantly different when measured in compression or in tension. For example, recent measurements of the $\pi$-coefficients in compression[9,21,22] are up to a factor of 2 larger than Smith's values in p-type material and almost half of the values obtained by Smith for n-type material. Full band *ab initio* calculations from large compressive stress to large tensile stress support these experimental conclusions[23].

**Section 3: Giant piezoresistance in Silicon nanostructures**

In semiconductor nanostructures that are sufficiently small, there can be a significant change in the electronic structure due to quantum confinement. Given that the PZR in bulk material is closely related to the electronic structure it is natural that it may also be modified by quantum confinement. The first experimental indication of this is the early work of Dorda in the surface inversion layer of bulk Silicon[24,25]. At low temperatures the PZR along particular crystal directions is found to be much larger than that of bulk Silicon with gage factors up to 1500. Both charge transfer and a stress-induced scattering mechanism resulting in a strongly stress-dependent $\tau_m$ in Eq. (4) have been invoked to describe these observations[26,27]. Quantum confinement leads to even more impressive, low temperature resistance changes in n-type AlAs quantum wells. In this case a 2-dimensional version of the charge transfer mechanism yields extraordinarily large gage factors up to 50000[28]. In p-type material, distortion of the hole band dispersion relationships results in similarly large gage factors[29].

Unlike these reports, there are as yet no experimental studies of PZR in semiconductor nanowires of sufficiently small diameter that quantum confinement begins to play a role. Work on Silicon nanowires started with Toriyama and co-workers, who studied the effect of applied stress first on polycrystalline nanowires[30] and then on top-down fabricated single crystal nanowires[31,32]. Although these authors report gage factors up to 50 % larger than Smith's results[5] in nanowires of approximately 50 nm diameter, there is nothing to suggest that the observed PZR is anything other than the known bulk phenomenon, particularly given the reported variations[33] in measured $\pi$-coefficients in bulk Silicon. Moreover, the applied stress was only roughly estimated using ex-situ finite element calculations of the nanowire geometry which may result in significant errors. The work which re-ignited interest in PZR in semiconductors was the report of giant PZR (GPZR) in bottom-up grown Silicon nanowires in 2006[18].

In bottom-up grown, single crystal <111> oriented Silicon nanowires subjected to both compressive and tensile stress, longitudinal $\pi$-coefficients up to 100 times larger than those of bulk Silicon were reported[18]. Not only is the PZR extremely large, but it is also strongly non-linear, with a large change of resistance per unit applied stress occurring for compressive stress rather than tensile stress, reminiscent of bulk Silicon PZR when studied at large stresses[9,21,22,23]. Several proposed explanations of the PZR phenomenon are based on quantum effects in atomically thin nanowires[34,35,36,37,38]. In one case at least[34], an interplay between the HH and LH electronic structure of the surface atomic layer and the interior atoms of the nanowire results in a predicted PZR behavior that strongly resembles the data of He and Yang. In other calculations a stress-induced shift from an indirect to a direct bandgap is predicted to yield a large sudden change in the resistance at several percent strain[35,37]. However, in some reports at small strains the calculated deformation potentials (i.e. the amount by which the electronic bands move per unit strain) in nanowires are very similar to those of bulk Silicon[8,9,37]. In these cases, the characteristic length scale for the predicted phenomena is the quantum, or de Broglie, wavelength. In Silicon at 300 K this is typically of the order of 1 nm which is much smaller than the diameters of the experimentally tested nanowires (which vary from approximately 50 nm up to 300 nm). As such, the nanowire diameters are too large for the GPZR to be any type of quantum size effect like those observed in surface accumulation layers or quantum wells[24,25,26,27,28,29]. These first principle calculations are therefore of interest for potential future PZR experiments on radially quantum confined nanowires[39] with diameters below approximately 2.2 nm[40]. In He and Yang's relatively large nanowires one therefore expects the electronic structure to be that of the bulk material, and it is for this reason that the apparent GPZR is such a surprising, unexpected and interesting result.

While He and Yang[18] did not attempt to describe the microscopic details of the PZR they observed, they naturally suspected that the surface had some role to play. It was shown for example that both the nanowire resistance and PZR depend strongly on surface chemical treatments; both are reduced when the nanowires are hydrogen passivated using a hydrofluoric acid etch. Re-oxidation using nitric acid makes both parameters increase. Moreover by studying the gate voltage dependence of the PZR they concluded that the GPZR was linked primarily to a change in the charge carrier mobility (and not the concentration), similar to the bulk PZR. A key realization concerning a possible origin for the GPZR effect came afterwards, when the correlation between partial depletion of the nanowire and GPZR was made[41].

By comparing the nanowire diameters with the characteristic electrostatic penetration depth – the surface depletion layer width – it was noticed that GPZR apparently occurs only in nanowires where the diameter is comparable to or smaller than the surface depletion layer width. When this is the case, the density of free charge carriers in the nanowire becomes sensitive to the charge state of the surface. For small enough diameters, the free carrier density within the conducting nanowire channel is entirely determined by the surface charge state. If this surface charge can be modified by an applied stress then a large change in carrier density (and hence conductivity) will result. This would at least qualitatively explain the sensitivity of the observed phenomena to surface chemical modification, but is at odds with He and Yang's claim that a mobility change is the principal origin of the GPZR. Despite this difference, the so-called piezopinch model[41] in which a stress-induced change of the surface charge density is modeled by a shift in the surface Fermi-level pinning, qualitatively describes the observed PZR as a function of nanowire diameter and doping level very well[41,42]. The only parameter in this model is the degree by which the surface Fermi-level shifts with applied stress, $dE_F/dX$. In order to match the He and Yang's observations $dE_F/dX \approx 0.5$ meV/MPa is found[41]. This is intriguingly close to stress-induced changes in trap activation energy measured in metal-oxide-Silicon capacitors[43], which lends some weight to the piezopinch idea. Unlike either the quantum mechanical confinement effects, or in the bulk effect, where the PZR is principally the result of a stress-induced change in the charge carrier mobility, the piezopinch model predicts a PZR based on a change in the charge carrier concentration[ii]. This is consistent with the known effects of surface state doping of nanowires[44]. He and Yang's data therefore tentatively support support an electrostatic interpretation of the GPZR whose characteristic length scale can be as large as several micrometers at low doping densities. Having said this, if the basic piezopinch idea is correct, the microscopic origin of the surface Fermi-level shift is yet to be identified.

At approximately the same time the piezopinch idea was developed, other authors had also noticed the fact that the apparent GPZR depends strongly on the doping density[45]. This early experimental work on top-down fabricated nanowires also reports large $\pi$-coefficients with a reduction in doping density for nanowires of fixed diameter, although the values are not quite as large as He and Yang's. In order to test the piezopinch and quantum confinement models, and to further explore the PZR in Silicon nanowires, a number of groups subsequently undertook experimental work on top-down fabricated nanowires[45,46,47,48,49,50], in sub-threshold nanowire transistor channels[51,52,53,54,55,56] and in Silicon thin films[48,57]. With some exceptions[58,59], very little subsequent work has been performed on bottom-up grown nanowires. Table 1 summarizes the measured gage factors and longitudinal $\pi$- coefficients along with the accepted values for bulk Silicon and those reported in Ref. 18. From the table, it is clear that the literature contains both claims and counter-claims of GPZR, and these will be discussed below.

---

[ii] A stress-induced change in the charge carrier concentration is also invoked to described the PZR of Silicon nano-needles under biaxial stress [see Z. Xiao, J. She, S. Deng and N. Xu, Appl. Phys. Lett. **110**, 114323 (2011)], although in this case a piezopinch description is not used.

| Type | *G* | $\pi$<br>$(\times 10^{-11}\ \mathrm{Pa}^{-1})$ | Ref. | Gated? | Stress<br>modulation? | Stress | Partial<br>Depletion? | Comments |
|---|---|---|---|---|---|---|---|---|
| bulk | | -102 [n, <100>]<br>+6.6 [p, <100>]<br>-31 [n, <110>]<br>+71.8 [p, <110>] | 5 | N | N | T | N | Light doping |
| bulk | | +25 [p, <110>] | 6 | N | N | T | N | Heavy doping |
| bulk | | -80 [n, <100>]<br>+150 [p, <110>]<br>+160 [p, <111>] | 9 | N | N | C | N | Light doping |
| top down | | +38.7 [p, <110>] | 32 | N | N | T | N | Released and non-released |
| top down | | +48 [p, <110>] | 31 | N | N | T | N | |
| top down | | +106 [p, <110>]<br>+455 [p, <110>] | 45,47 | N | N | C | N<br>Y | Dry oxide |
| top down | | +82 [p, <110>]<br>+130 [p, <110>] | 46 | N | N | T | N<br>Y? | Dry oxide |
| top down | | +65 [p, <110>]<br>-70 [n, <110>] | 48 | N | Y | C/T | Y | Non-released |
| top down | | +96 [p, <110>]<br>+120 [p, <110>]<br>+205 [p, <110>] | 48 | N | Y | C/T | Y | Released |
| top | 350 | +206 [p, <110>] | 50 | N | N | C | Y | R drift, higher |
| down | | | | | | | | PZR on surfaces with higher trap densities |
| top<br>down | | +1538 [p, <110>] | 52 | Y | N | T | Y | Sub-threshold effect |
| top<br>down | 5000 | *+2940* [p, <110>] | 56 | Y | Y | C/T | Y | Sub-threshold effect |
| top<br>down | 25<br>50<br>250 | *+14.7* [p, <110>]<br>*+29.4* [p, <110>]<br>*+147* [p, <110>] | 49 | N | N* | C/T | Y | Heavy doping<br>Med. doping<br>Released |
| top<br>down | | -400 [p, <110>]<br>-400 [trans.] | 57 | N | N | C/T | Y | + transverse π<br>same as<br>longitudinal<br>Thin film |
| top<br>down | 2500 | *+1470* [n, <110>] | 53,54 | Y | Y | T | Y | Sub-threshold effect |
| top<br>down | | +207 [n, n/s] | 55 | Y | N | C/T | Y | Sub-threshold effect |
| top<br>down | | +25 [p, <110>]<br>+190 [p, <110>] | [60] | N | N | T high | N<br>Y | Released |
| bottom<br>up | | +3550 [p, <111>]<br>+3100 [p, <110>] | 18 | N/Y[§] | N | C/T | Y | Released |
| bottom<br>up | <br>-250 | +25 [p, <111>]<br>-131 [p, <111>] | 58 | N | N | T low<br>T high | Y | Released |
| bottom | -460 | *-219* [p, <111>] | 59 | N | N | T high | Y | Released |
| up | | | | | | | | Largest effect at negative bias. p.59 thesis |

Table 1: Summary of longitudinal $\pi$-coefficients and gage factors ($G$) reported in bulk and nanostructured Silicon. In cases where only $G$ is given, the Young's modulus for the given crystal direction[61] is used to calculate the $\pi$-coefficient (these values being given in italics). *Resistance variations are carefully monitored as a function of time. Reported results are obtained during periods where the zero-stress resistance does not vary significantly with time. §In He and Yang's work the gated nanowires are also tested with the aim of determining whether the PZR is due to a mobility or concentration change. The effect on the PZR of placing the transistor structure in the sub-threshold regime is not specifically discussed.

**Subsection 3.1: Effect of electrostatic depletion**

Electrostatic depletion in nanowires or thin films is obtained in un-gated structures by reducing the doping concentration (which in the most commonly used p-type material is denoted $N_A$) so that the surface depletion layer width,

$$W = \sqrt{\frac{2\varepsilon\phi_b}{qN_A}}, \qquad (7)$$

increases to the point where it is comparable to (or larger than) the smallest dimension of the nanowire or thin film, $L$. Here $\varepsilon$ is the dielectric constant of the material, $q$ the absolute value of the electronic charge and $\phi_b$ is the surface potential barrier resulting from the presence of charge trapping in localized surface states[62]. The actual value of $\phi_b$ depends on the microscopic details of the surface states themselves, the doping in the bulk of the semiconductor and the crystal orientation of the surface[63]. In Silicon, it can vary over more than 0.7 eV relative to the bulk position of the Fermi level[64,65,66]. This will clearly affect the critical doping density at which $W \sim L$. It is reasonable however, for an order-of-magnitude estimate of the depletion of nanowires and thin films, and in order to simplify the following discussion, to assume that the $\phi_b$ = 0.55 eV i.e. that the surface Fermi level is pinned in the middle of the bandgap. In the cases where partial depletion is induced by reducing the doping density (and not by depleting a transistor channel with a gate voltage), an estimation of $W$ is made in this way for the literature summarized in Table 1. Figure 4 shows a plot of the reported longitudinal $\pi$-coefficients versus the ratio $W/L$.

This figure is particularly informative concerning the state of the art in Silicon nanowire PZR. The two horizontal, black lines represent typical minimum and maximum values of the bulk p-type longitudinal $\pi$-coefficients. Although there are crystal directions for which the PZR response is zero[10], the minimum value taken here is that measured in highly doped material along the <110> direction[6]. The maximum value is that obtained under compression in lightly doped material along the same crystal direction[9]. Although not all measured nanowires are orientated with their long axes along the <110> direction, for the majority this is the case. A striking feature of Fig. 4 is the relative lack of reports of GPZR with $\pi$-coefficients significantly outside the accepted range of reported bulk values. Apart from He and Yang's report[18], the only other claims of significantly larger PZR are those of Reck et al.[45,47], and Yang and Li[57]. Koumela et al. reports values slightly above the acceptable bulk range[49].

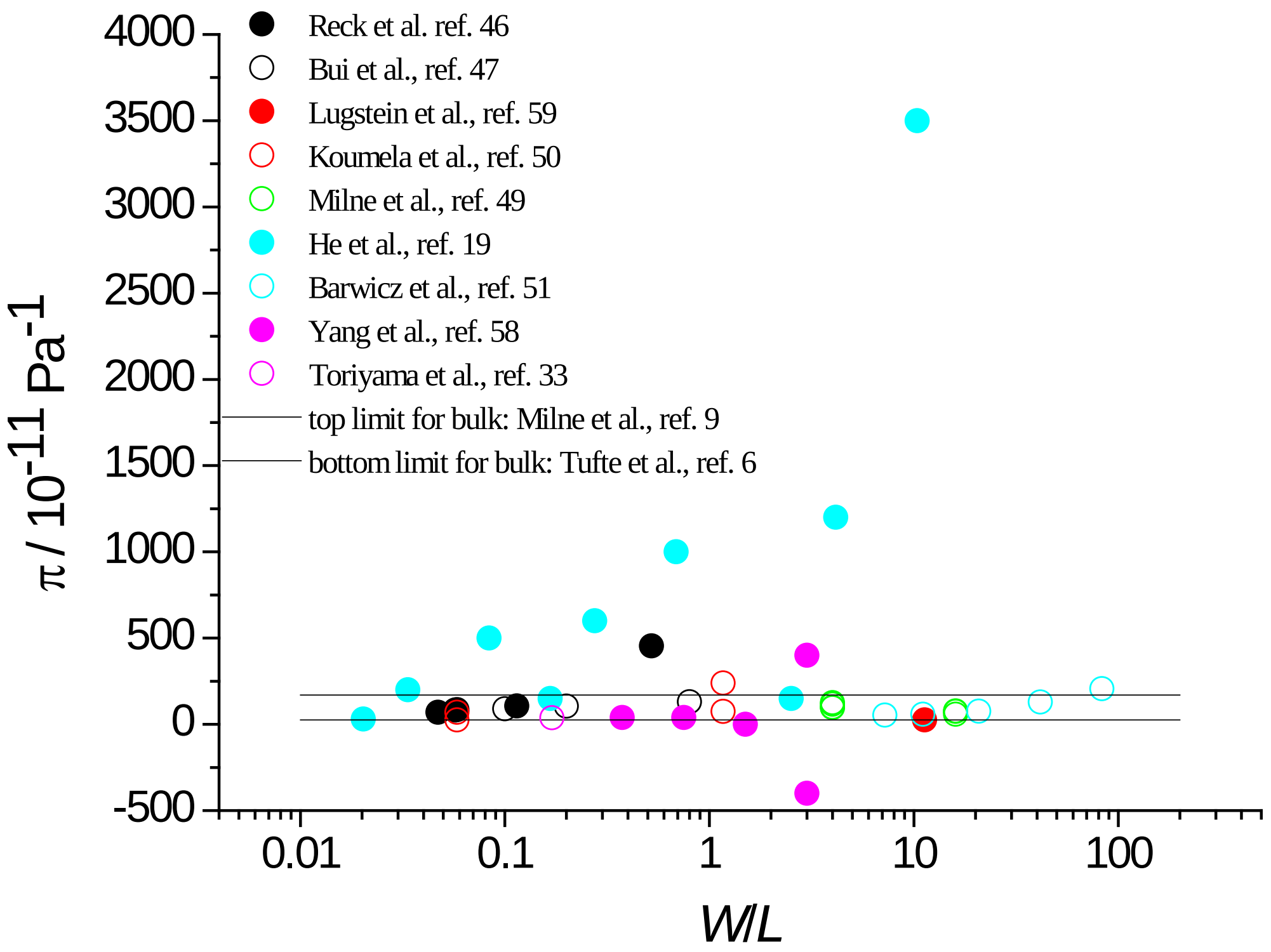

Figure 4: Summary of measured longitudinal $\pi$-coefficients in p-type Silicon nanowires and thin films as a function of carrier depletion as measured by the ratio *W*/*L* (see text). The higher the value of *W*/*L*, the more the structure is depleted of free charge carriers. The horizontal black lines represent minimum and maximum experimentally measured values in bulk, p-type Silicon.

Reck's work[45] is intriguing since not only does it report large PZR, but (as mentioned above) it also explores the variation in PZR with doping density. It is clear from Fig. 4 that these authors measure a progressive increase in the longitudinal $\pi$-coefficient with increasing *W*/*L*. While this is the tendency that is expected based on quantum degeneracy arguments in bulk Silicon[10], the maximum values that are measured appear to be significantly larger than the largest reported values in bulk Silicon. Contrast this with another interesting work on nanowires where the surface crystal orientation is found to affect the PZR[50]. In this case the same tendency with a change in doping density is observed, but the measured $\pi$-coefficients fall much closer to the known range of bulk values (see also Fig. 4). It is therefore tempting to suggest that Reck's reported result is evidence of a phenomenon different to that of bulk Silicon, and one might tentatively conclude that a piezopinch description[41,42] would suitably describe the data. Indeed, the $\pi$-coefficient values shown in Fig. 4 and in Table 1 are not far from those of He and Yang[18] for equivalent *W*/*L* ratios.

Another report of GPZR is that of Yang and Li in very thin Silicon films[57]. In this case the authors go to some length to point out that a quantum mechanical confinement effect is not the cause of the observed effect, and that a large effect is observable only for the thinnest films where $W/L >> 1$. Along the lines of the piezopinch model, they suggest that the explanation for the GPZR is a stress-induced change in the surface charge density. They make two intriguing experimental claims, i) that the longitudinal PZR coefficient is large and identical (in magnitude and sign) to the transverse coefficient. This clearly indicates that if the PZR effect is real, it is *not* a bulk electronic structure effect and ii) that the longitudinal $\pi$-coefficient changes sign with a change from tensile to compressive applied stress. i.e. that regardless of the sign of the applied stress, the resistance always decreases (see Figure 3(c) in Ref. 57). The second of these claims is rather difficult to justify in terms of a real PZR effect. An even response as a function of stress suggests that the Fermi energy at zero-stress sits exactly on a degenerate pair of identical states whose stress response (shift in energy, change in curvature etc…) are exactly equal and opposite. There are no known Silicon surface states that possess this property[63,64,65]. A more likely explanation for the observed resistance change is simply a non-stress-related surface charge relaxation[48,67]. With reference to Figure 3(c) in Ref. 57, it is likely that the true PZR is the slight difference in the slopes of the curves for tensile and compressive applied stress. This is possibly just the bulk Silicon PZR.

This charging effect, or dielectric relaxation, was first revealed in nanowires in the study by Milne et al., where the true PZR of a large set of top-down nanowires was found to be comparable to that of bulk Silicon[48]. Milne went further to suggest that the non-stress-related surface charging effects, which only show up in partially depleted structures, might well lead to claims of apparent GPZR. While Yang and Li's work appears to fall into this catergory[57], the question with respect to the data shown in Fig. 4, is whether He's[18], or indeed Reck's[45,47], does as well. He's work on the effect of surface functionalization on nanowire resistance and PZR clearly shows that the observed phenomena are sensitive to surface charge, but there is no outright published evidence that the reported resistance changes are due to anything other than the applied stress. This is of course far from being proof that the large resistance changes are due to the applied stress. The same is true of Reck's work. Without further investigation, it is difficult to argue either for the unambiguous existence of GPZR in nanowires or for clear signs of dielectric relaxation. Having said this, Milne[48] showed that it is simple to design a time-modulated stress experiment on samples like He's or Reck's that would greatly clarify the interpretation of these two early reports of GPZR. While this technique has been employed by several authors who investigate gated or transistor-like nanowires[53,54,56], it has not been widely employed in un-gated (i.e. air exposed) nanostructures.

In their work, Koumela et al.[49] were careful to monitor resistances as a function of time in order to avoid any possible misinterpretation. They pursued a slightly different route to others and chose to investigate the effect of releasing the nanowires from the buried oxide (BOX). Their observations suggest that releasing nanowires systematically increases the $\pi$-coefficient. For example, a nanowire with a Boron doping density of 2 x $10^{20}$ $cm^{-3}$ ($W/L$ ~ 0.06) has a measured longitudinal $\pi$-coefficient of approximately 25 x $10^{-11}$ $Pa^{-1}$ in good agreement with bulk material[6]. When the 145 nm thick BOX is removed this increases to approximately 60 x $10^{-11}$ $Pa^{-1}$, almost a factor of 3 increase. Similarly, a less doped nanowire (5 x $10^{17}$ $cm^{-3}$, $W/L$ ~ 1.2) sees its $\pi$-coefficient increase from 70 x $10^{-11}$ $Pa^{-1}$ to 240 x $10^{-11}$ $Pa^{-1}$, a boost of just over 3. It seems unlikely that the nature of the surface oxide has changed during the release process – both before and afterwards it is just the native oxide. On the other hand, the actual diameter of the Silicon nanowire itself is probably slightly reduced from its pre-release value due to the etch process. This would thereby increase $W/L$ and perhaps lead to an increase in the $\pi$-coefficient according to the piezopinch model[41]. Another possibility is a stress concentration effect[68,69] due to the geometry of the released nanostructure. In this case, the real stress along the length of the nanowire may be significantly larger than the externally applied stress with the measured resistance change being due simply to the bulk Silicon PZR. It is also interesting to note that Milne[48] measured boosts in $\pi$-coefficients up to a factor of 2 in nanowires upon their release (and without any other change in the nanowire itself). Such a stress concentration phenomenon may affect, to a lesser extent, un-released nanowires.

The final important feature of Fig. 4 that should be discussed is the relative lack of work performed on bottom-up grown nanowires – in this figure there are only two examples, that of He and Yang[18] and the work of Lugstein et al.[58]. Lugstein's results are rather different to those of He and Yang, with only small $\pi$-coefficients being measured despite very strong depletion of the nanowires. Intriguingly, Lugstein uses the stress modulation technique[48] in order to avoid resistance drifts that might be interpreted as a GPZR effect. At high stresses in the GPa range, the $\pi$-coefficients of these <111> oriented change sign, and this observation has been confirmed elsewhere[59]. This in contrast to Kumar Bhaskar's work on <110>

oriented, top-down nanowires at GPa stresses where the $\pi$-coefficients do not change sign[60]. While the results on bottom-up grown nanowires are therefore inconsistent, the fact that the biggest values in Fig. 4 are claimed in this type of wire may not be without merit. Having said this, *if* there is a significant difference between bottom-up and top-down fabricated material, it is yet to be identified, although it is possible to speculate on a number of potentially important differences.

Firstly, since the phenomena appear to be related to surface states, the fact that bottom up wires have facetted surfaces may be important since this is known to affect the surface state distribution[63]. This might differentiate the properties of He and Yang's nanowires which have facets normal to the {211} crystal directions, from Lugstein's nanowires which have {111} oriented surface facets. Depending on the exact microscopic details of the atomic configurations associated with a particular trap state, the state's energy may be stress dependent[43] or not. Secondly, bottom-up grown nanowires are known to contain significant quantities of the catalytic metal, usually gold, in the body of the nanowires as well as on the surface[70]. Given that Gold is known to form deep charge traps in Silicon[71], if the activation energy of these localized states is found to be stress-dependent, they might be responsible for a piezopinch effect in partially-depleted, bottom-up nanowires that would be absent in top-down nanowires. Another potentially important point is the surface roughness of the nanowire, which tends to be greater in dry-etched, top-down structures. Reck[45] points out that surface roughness scattering could conceivably quench any piezoresistive effect, including the bulk one. In this picture, even if GPZR exists in top-down nanowires, it may be quenched by this scattering process.

**Subsection 3.2 Gated or transistor-like nanowires**

Table 1 tells a rather different story in gated or transistors-like nanowires in which a gate potential can be used to place the transistor in its sub-threshold regime. In this case there are several claims of GPZR with $\pi$-coefficients well above the bulk range indicated in Fig. 4. These will be individually examined in this section.

The first such work is that of Passi et al.[52], who find that the longitudinal $\pi$-coefficient varies from an acceptable bulk value in the linear response region of the transfer curve up to 1600 x $10^{-11}$ $Pa^{-1}$ in the sub-threshold region. Although the stress-modulation method is not used, the reproducibility of the results as a function of gate bias for several different applied stresses suggests that there is no dielectric relaxation causing resistance drift[48,67]. Subsequent work has usually attempted to address this possible shortcoming directly by implementing some form of dynamically applied stress[53,54,55,56]. In all of these cases, GPZR in the sub-threshold region is observed, with the largest claimed values[56] approaching those measured by He and Yang on un-gated bottom-up nanowires[18].

In Neuzil's case[56], the geometry of the sample is rather unusual, with two parallel nanowires connected electrically in series. Moreover, the wires are placed in a moveable cantilever which is actuated in order to apply stress. The gate is the fixed substrate itself. An unusual feature of the transfer curves of the nanowires is the sub-threshold slope (i.e. $dI_{DS}/dV_G$) which is approximately 160 mV/decade. This is comparable with the best performance of recently developed tri-gate, 22 nm node devices fabricated by Intel[72], which is rather surprising for a relatively large, un-optimized device structure in which the gate electrode is some distance from the conducting nanowire channel itself. This behavior suggests that there is some geometric boost to the effective gating effect that may itself be of interest. One possibility might be the close proximity of the two parallel nanowires that yields some electrostatic coupling between the two. Whether this apparently very large sub-threshold slope has an effect on the PZR is not clear, but the electrical behavior of this sample geometry is clearly more complex than a single isolated, gated nanowire geometry.

Such criticism cannot be leveled at Singh's[55], and in particular at Kang's work[53,54]. In these cases the sub-threshold slopes are much more modest as would be expected for devices of this size. Like Passi's work[52], very large, clear-cut changes in the nanowire resistance under applied stress are observed in the sub-threshold region. In these cases, the giant PZR seems therefore to be reproducible, and the question is then: why are gated structures so different from un-gated ones where there is as yet no convincingly repeated evidence of GPZR?

The fact that the GPZR occurs only in the sub-threshold region of gated structures certainly suggests an electrostatic origin for the effect i.e. a stress-induced effective gating of the nanowire channel. Two possibilities present themselves – either a stress-driven effect where the applied stress modifies the density of trapped charge in the gate oxide, or a strain-driven effect where the applied stress modifies the distance between the gate electrode and the nanowire channel. This first of these is just the piezopinch effect[41], and it closely resembles early device ideas in which the gate oxide of transistors is replaced with a piezoelectric dielectric in order to render the transistor highly sensitive to applied stress[73]. In this case gage factors up to 7000 are observed. It is however difficult to understand why a piezopinch mechanism would not also be clearly present in un-gated, oxidized nanowires. The second possibility seems unlikely given that in the majority of works[52,53,54,55], the sub-threshold slope is not large enough to account for such a large resistance change under such a small resulting strain i.e. the physical motion of the gate electrode relative to the channel due to the applied stress is probably too small or even negligible. Having said this, in a non-uniform object like a gated nanowire, stress concentration[68,69] may result in much larger-than-expected strains so this possibility should not be ruled out without further investigation.

**Section 3.3 Silicon nanowire strain gages**

There are a number of publications in the literature whose goal is to use Silicon nanowire elements as strain gages for the detection of a variety of micro-electromechanical (MEMS) or nano-electromechanical (NEMS) transduced forces[74,75,76,77,78,79]. In some cases[75,79] the authors recognize that while GPZR in partially depleted nanowires would be attractive from a force sensitivity viewpoint, the partial depletion leads to very high resistances and to an augmented sensitivity to environmental changes (i.e. humidity, temperature, etc…) resulting in dielectric relaxation[48]. In the view of these authors, exploitation of any nanowire PZR is therefore best achieved by using high doping densities to reduce resistance and surface sensitivity while maintaining a PZR effect based on the known bulk phenomenon at high charge carrier density[6,10]. Other articles either directly claim[74] to exploit the GPZR, or simply imply its use[77] in a selected application. In these works it is very difficult to determine whether the resulting device sensitivity is a consequence of GPZR, or whether it is simply a combination of bulk Silicon PZR and a well-designed detection protocol. As such, they do not add significantly to the discussion of the GPZR effect itself.

Still other nanowire strain gage applications use a gated nanowire geometry[76,78]. Given the discussion above in Section 3.2, in this geometry any increase in sensitivity is potentially the result of GPZR. One case[76] uses the same unusual double-nanowire geometry employed by Neuzil[56], although ultra-high sub-threshold slopes are not reported. This work by Singh et al.[76] is intriguing since it is the only report of both the gate effect and a reduction in doping density on nanowire PZR. Their data supports the main observations given in this review – a reduction in doping density increases the $\pi$-coefficient, and this increase is much more pronounced when the nanowire transistor is placed in its sub-threshold region. Sansa's work[78] is also interesting in that it employs a separate gate and nanowire structure that in principal permits a stress-induced movement of the gate with respect to the nanowire. As indicated in section 3.2, if such a movement occurs when the transistor is in the sub-threshold region, the sub-threshold slope will determine the magnitude of the PZR.

**Section 3.4 Other potentially important issues**

There are several other important issues which could potentially complicate the interpretation of PZR data in nanowires. The two most significant of these are i) in-built stress within the nanowire, either as grown for bottom-up nanowires, or residual stress in the active layer of the Silicon-on-insulator (SOI) films in the case of top down nanowires, and ii) poorly characterized electrical contacts, in particular to partially depleted bottom-up grown nanowires.

In the case of the residual stress in SOI thin films, the active layer stress resulting from the fabrication procedure has been known for some time[80]. Recently, in-built compressive stresses up to more than 300 MPa were found in commercial SOI wafers[81]. This level of stress is clearly not negligible when compared with the typical external stresses applied to nanowires during PZR measurements which are usually at most approximately 100 MPa. Moreover, since the PZR response of bulk Silicon is strongly non-linear, even around zero applied stress[9,23], an in-built stress might significantly modify the apparent $\pi$-coefficient of top down nanowires fabricated from SOI wafers, even if the PZR response is nothing more than that of bulk Silicon. Similarly, it is entirely possible that the growth of bottom-up grown nanowires between confining side walls[18], the evaporation of electrical contacts onto nanowire bridges[58], or the use of external probes to measure nanowire resistance[59], may also result in an uncontrolled, offset stress in these materials[iii]. These issues are potentially important because the PZR response to an external applied stress can depend on the initial stress state of the nanowire[51].

The lack of knowledge about the electrical properties of the nominally Ohmic contacts to partially depleted nanowires is also problematic. In principal in top down nanowires, even partially depleted ones, low resistance Ohmic contacts can be formed in the contiguous active Silicon layer by the usual implantation and metal evaporation techniques[48]. On the other hand, in bottom-up grown nanowires there is not only likely to be a significant quantity of metal catalyst at one or both ends of the nanowire, but the formation of a space charge junction between the lightly doped nanowire and a more heavily doped substrate cannot be ruled out. In this case, it is possible that a rectifying junction is formed at each end of the nanowire so that the overall object's equivalent circuit is given by back-to-back diodes[82]. This could also influence the PZR properties of the nanowires[83].

[iii] Given that the state of the residual stress is not usually characterized, its probable presence justifies the comparison of PZR data under externally applied compressive and tensile loads in Table 1 despite the expected differences between the two (see Section 2).

**Section 4: Conclusions and future possible research directions**

The literature strongly indicates that GPZR does exist in partially depleted Silicon nanowires, but only in structures containing an electrostatic gate plane that biases the nanowire into the sub-threshold region. Unambiguous proof of GPZR in un-gated structures is lacking, with the majority of works reporting $\pi$-coefficients that are conceivably linked to a bulk Silicon PZR phenomenon, or that are potentially marred by the effects of dielectric relaxation. A piezopinch (or electrostatic) explanation may be the correct qualitative interpretation for the GPZR in transistor-like structures, but this is incompatible with the absence of a clear effect in un-gated nanowires. Possible future research directions to address this issue might include i) a study of the stress-dependence of trap state activation energies in thermal and native oxides, for example by deep level transient spectroscopy or by using Kelvin probe microscopy. This might help to address the microscopic origins of an effect based on the piezopinch idea, ii) an investigation of the effects of the stress distribution in the nanostructures, either with an eye to estimating the effect of stress concentration or to evaluate the possibility of a change in gate/channel distance, iii) a more thorough understanding of in-built stresses in the nanowires, and iv) an evaluation of the electrical properties of the contacts, especially in the case of partially depleted, bottom-up nanowires where the possibility for the formation of rectifying junctions at each end of the wire exists.

A second axis for future research concerns the exploration of the PZR properties of bottom-up nanowires. There is very little literature on this type of nanostructure, and given that the biggest $\pi$-coefficients are claimed in bottom-up materials more work is certainly justified. Potentially important differences between top-down and bottom-up nanowires could then be explored, for example a potential role for Gold-related trap states in the piezopinch model. The principal challenge in this case is the fabrication of Ohmically contacted nanowires which is significantly more difficult than in the case of top-down nanowires and nanostructures. A third possibility is to push nanowire dimensions down to quantum length scales where a number of interesting phenomena, including GPZR, are theoretically predicted based on *ab initio* band structure calculations. This is currently a challenge for growers of bottom-up Silicon nanowires where easily achievable minimum nanowire diameters are of the order of 30 nm.

GPZR is the basis for an enabling technology for a wide range of force transduction applications which require the sensitive electrical detection of MEMS or NEMS motion. Moreover, mechanical stress is a key ingredient in the semiconductor technology roadmap for the improvement of microelectronic device performance. Aside from the purely scientific goal of understanding the phenomenon, a better understanding of the PZR properties of nanowires and nanowire transistors should therefore be a priority.

**Acknowledgements**

This work was supported by the Agence Nationale de la Recherche (PIGE ANR-2010-021). The author would like to thank S. Arscott for many fruitful discussions and for a critical reading of the manuscript.

## REFERENCES

1. P. W. Bridgman: The electrical resistance of metals under pressure. Proc. Am. Acad. Arts & Sci. **52**, 573 (1917)

2. P. W. Bridgman: The resistance of 72 Elements, alloys and compounds to 100,000 Kg/Cm², Proc. Am. Acad. Arts & Sci. **81**, 165 (1952)

3. L. Fu, C. L. Kane and E. J. Mele: Topological insulators in three dimensions, Phys. Rev. Lett. **98**, 106803 (2007)

4. J. C. Wolfe: Summary of the Kronig-Penney electron, Am. J. Phys. **46**, 1012 (1978)

5. C. S. Smith: Piezoresistance effect in germanium and silicon, Phys. Rev. **94**, 42 (1954)

6. O. N. Tufte and E. L. Stelzer: Piezoresistive properties of heavily doped n-type silicon, J. Appl. Phys. **34**, 313 (1963)

7. C. Herring and E. Vogt: Transport and deformation-potential theory for many-valley semiconductors with anisotropic scattering, Phys. Rev. **101**, 944 (1956)

8. J. Aubrey, W. Gubler, T. Henningsen, and S. Koenig: Piezoresistance and Piezo-Hall-Effect in n-Type Silicon, Phys. Rev. **130**, 1667 (1963)

9. J.S. Milne, I. Favorskiy, A. C. H. Rowe, S. Arscott, Ch. Renner : Piezoresistance in silicon at uniaxial compressive stresses up to 3 GPa, Phys. Rev. Lett. **108**, 256801 (2012)

10. Y. Kanda : A graphical representation of the piezoresistance coefficients in silicon, I.E.E.E. Trans. Elec. Dev. **29**, 64 (1982)

11. E. Adams: Elastoresistance in p-type Ge and Si, Phys. Rev. **96**, 803 (1954)

12. Y. Ohmura: Piezoresistance effect in p-type Si, Phys. Rev. B **42**, 9178 (1990)

13. K. Suzuki, H. Hasegawa, and Y. Kanda: Origin of the linear and nonlinear piezoresistance effects in p-type silicon, Jpn. J. Appl. Phys. **23**, L871 (1984)

14. P. Kleimann, B. Semmache, M. Le Berre, and D. Barbier: Stress-dependent hole effective masses and piezoresistive properties of p-type monocrystalline and polycrystalline silicon, Phys. Rev. B **57**, 8966 (1998)

15. J. Richter, J. Pedersen, M. Brandbyge, E. Thomsen, and O. Hansen: Piezoresistance in p-type silicon revisited, J. Appl. Phys. **104**, 023715 (2008)

16. S. Thompson, G. Sun, Y. Choi, and T. Nishida: Uniaxial-process-induced strained-Si: extending the CMOS roadmap, I.E.E.E. Trans. Electron Devices **53**, 1010 (2006)

17. X. Fan, L. Register, B. Winstead, M. Foisy, W. Chen, X. Zheng, B. Ghosh, and S. Banerjee: Hole mobility and thermal velocity enhancement for uniaxial stress in Si up to 4 GPa, I.E.E.E. Trans. Electron Devices **54**, 291 (2007)

18. R. He and P. Yang: Giant piezoresistance effect in silicon nanowires, Nature Nanotech. **1**, 42 (2006)

19. K. Matsuda, Y. Kanda, K. Yamamura and K. Suzuki: Second-order piezoresistance coefficients of p-type silicon, Jap. J. Appl. Phys. **29**, L1941 (1990)

20. K. Matsuda, K. Suzuki, K. Yamamura and Y. Kanda: Nonlinear piezoresistance effects in silicon, J. Appl. Phys. **73**, 1838 (1993)

21. L. Shifren, X.Wang, P. Matagne, B. Obradovic, C. Auth, S. Cea, T. Ghani, J. He, T. Hoffman, R. Kotlyar, Z. Ma, K. Mistry, R. Nagisetty, R. Shaheed, M. Stettler, C. Weber, M. D. Giles: Drive current enhancement in p-type metal–oxide–semiconductor field-effect transistors under shear uniaxial stress, Appl. Phys. Lett. **85**, 6188 (2004)

22. Y. Tsang, A. O'Neill, B. Gallacher, and S. Olsen: Using piezoresistance model with cr conversion for modeling of strain-induced mobility, I.E.E.E. Elec. Dev. Lett. **29**, 1062 (2008)

23. S. I. Kozlovskiy and N. N. Sharan: Piezoresistive effect in p-type silicon classical nanowires at high uniaxial strains, J. Comput. Electron. **10**, 258 (2011)

24. G. Dorda : Effective mass change of electrons in silicon inversion layers observed by piezoresistance, Appl. Phys. Lett. **17**, 406 (1970)

25. G. Dorda : Piezoresistance in quantized conduction bands in silicon inversion layers, J. Appl. Phys. **42**, 2053 (1971)

26. I. Eisele: Stress and intersubband correlation in the silicon inversion layer, Surf. Sci. **73**, 315 (1978)

27. G. Dorda, I. Eisele and H. Gesch: Many-valley interactions in n-type silicon inversion layers, Phys. Rev. B **17**, 1785 (1978)

28. Y. P. Shkolnikov, K. Vakili, E. P. De Poortere, M. Shayegan: Giant low-temperature piezoresistance effect in AlAs two-dimensional electrons, Appl. Phys. Lett. **85**, 3766 (2004)

29. B. Habib, J. Shabani, E. P. De Poortere, M. Shayegan, and R. Winkler: Anisotropic low-temperature piezoresistance in (311)A GaAs two-dimensional holes, Appl. Phys. Lett. **91**, 012107 (2007)

30. T. Yasutada, T. Toriyama, and S. Sugiyama: Characteristics of polycrystalline Si nano wire piezoresistors, Tech. Digest Sens. Symp. **17**, 195 (1999)

31. T. Toriyama, Y. Tanimoto and S. Sugiyama: Single crystal silicon nano-wire piezoresistors for mechanical sensors, J. Mems Sys. **11**, 605 (2002)

32. T. Toriyama, D. Funai and S. Sugiyama: Piezoresistance measurement on single crystal silicon nanowires, J. Appl. Phys. **93**, 561 (2003)

33. R. E. Beaty, R. C. Jaeger, J. C. Suhling, R. W. Johnson and R. D. Butler: Evaluation of piezoresistive coefficient variation in silicon stress sensors using a four-point bending test fixture, I.E.E.E. Trans. Comp., Hyb., Man. Tech. **15**, 904 (1992)

34. J. X. Cao, X. G. Gong and R. Q. Wu: Giant piezoresistance and its origin in Si (111) nanowires: First-principles calculations, Phys. Rev. B **75**, 233302 (2007)

35. D. Shiri, Y. Kong, A. Buin and M. P. Anantram: Strain induced change of bandgap and effective mass in silicon nanowires, Appl. Phys. Lett. **93**, 073114 (2008)

36. K. Nakamura, D. V. Dao, B. T. Tung, T. Toriyama and S. Sugiyama: Piezoresistive effect in silicon nanowires—a comprehensive analysis based on first-principles calculations, I.E.E.E. Conf. Micro-nanomech. 38 (2009)

37. P. W. Leu, A. Svizhenko and K. Cho: Ab initio calculations of the mechanical and electronic properties of strained Si nanowires, Phys. Rev. B **77**, 235305 (2008)

38. Y.-M. Niquet, C. Delerue and C. Krzeminski: Effects of strain on the carrier mobility in silicon nanowires, Nano Lett. **12**, 3545 (2012)

39. L. T. Canham: Silicon quantum wire array fabrication by electrochemical and chemical dissolution of wafers, Appl. Phys. Lett. **57**, 1046 (1990)

40. X. Zhao, C. M. Wei, L. Yang and M. Y. Chou: Quantum confinement and electronic properties of silicon nanowires, Phys. Rev. Lett. **92**, 236805 (2004)

41. A. C. H. Rowe: Silicon nanowires feel the pinch, Nat. Nanotech. **3**, 312 (2008)

42. T. T. Nghiem, V. Aubry-Fortuna, C. Chassat, A. Bosseboeuf and P. Dollfus: Monte Carlo simulation of giant piezoresistance effect in p-type silicon nanostructures, Mod. Phys. Lett. B **25**, 995 (2011)

43. A. Hamada and E. Takeda: Hot-electron trapping activation energy in PMOSFET's under mechanical stress, I.E.E.E. Elec. Dev. Lett. **15**, 31 (1994)

44. U. Kumar Bhaskar, T. Pardoen, V. Passi and J.-P. Raskin: Surface states and conductivity of silicon nano-wires, J. Appl. Phys. **113**, 134502 (2013)

45. K. Reck, J. Richter, O. Hansen and E. V. Thomsen: Piezoresistive effect in top-down fabricated silicon nanowires, I.E.E.E. Conf. MEMS, 217 (2008)

46. T. T. Bui, D. V. Dao, K. Nakamura, T. Toriyama and S. Sugiyama: Evaluation of the piezoresistive effect in single crystalline silicon nanowires, I.E.E.E. Sensors Conf. 41 (2009)

47. K. Reck, J. Richter, O. Hansen and E. V. Thomsen: Increased piezoresistive effect in crystalline and polycrystalline Si nanowires, Nanotechnology 920 (2008)

48. J. S. Milne, A. C. H. Rowe, S. Arscott and Ch. Renner: Giant piezoresistance effects in silicon nanowires and microwires, Phys. Rev. Lett. **105**, 226802 (2010)

49. A. Koumela, D. Mercier, C. Dupré, G. Jourdan, C. Marcoux, E. Ollier, S. T. Purcell and L. Duraffourg: Piezoresistance of top-down suspended Si nanowires, Nanotechnology **22**, 395701 (2011)

50. T. Barwicz, L. Klein, S. J. Koester and H. Hamann: Silicon nanowire piezoresistance: Impact of surface crystallographic orientation, Appl. Phys. Lett. **97**, 023110 (2010)

51. F. Rochette, M. Cassé, M. Mouis, A. Haziot, T. Pioger, G. Ghibaudo and F. Boulanger: Piezoresistance effect of strained and unstrained fully-depleted silicon-on-insulator MOSFETs integrating a $HfO_2$/TiN gate stack, Sold State Comms. **53**, 392 (2009)

52. V. Passi, F. Ravaux, E. Dubois and J. P. Raskin : Backgate bias and stress level impact on giant piezoresistance effect in thin silicon films and nanowires, I.E.E.E. Conf. MEMS, 464 (2010)

53. T. K. Kang: The piezoresistive effect in n-type junctionless silicon nanowire transistors, Nanotechnology **23**, 475203 (2012)

[54] T. K. Kang: Evidence for giant piezoresistance effect in n-type silicon nanowire field-effect transistors, Appl. Phys. Lett. **100**, 163501 (2012)

[55] P. Singh, W. T. Park, J. Miao, L. Shao, R. Krishna Kotlanka and D. L. Kwong: Tunable piezoresistance and noise in gate-all-around nanowire field-effect-transistor, Appl. Phys. Lett. **100**, 063106 (2012)

[56] P. Neuzil, C.C. Wong and J. Reboud: Electrically controlled giant piezoresistance in silicon nanowires, Nano Lett. **10**, 1248 (2010)

[57] Y. Yang and X. Li: Giant piezoresistance of p-type nano-thick silicon induced by interface electron trapping instead of 2D quantum confinement, Nanotechnology **22**, 015501 (2011)

[58] A. Lugstein, M. Steinmair, A. Steiger, H. Kosina and E. Bertagnolli: Anomalous piezoresistance effect in ultrastrained silicon nanowires, Nano Lett. **10**, 3204 (2010)

[59] Y. Zhang, X. Y. Liu, C. H. Ru, Y. L. Zhang, L. X. Dong and Y. Sun Y: Piezoresistivity characterization of synthetic silicon nanowires using a MEMS device, IEEE/ASME J. Microelectromech. Syst. **20**, 959 (2011)

[60] U. Kumar Bhaskar, T. Pardoen, V. Passi and J.-P. Raskin: Piezoresistance of nano-scale silicon up to 2 GPa in tension, Appl. Phys. Lett. **102**, 031911 (2013)

[61] J. J. Wortman and R. A. Evans: Young's modulus, shear modulus, and Poisson's ratio in silicon and germanium, J. Appl. Phys. **36**, 153 (1965)

[62] D. Vu, S. Arscott, E. Peytavit, R. Ramdani, E. Gil, Y. André, S. Bansropun, B. Gérard, A. C.H . Rowe and D. Paget: Photoassisted tunneling from free-standing GaAs thin films into metallic surfaces, Phys. Rev. B, **82**, 115331 (2010)

[63] F. J. Himpsel, G. Hollinger and R. A. Pollak: Determination of the Fermi-level pinning position at Si (111) surfaces, Phys. Rev. B **28**, 7014 (1983)

[64] L. M. Terman: An investigation of surface states at a silicon/silicon oxide interface employing metal-oxide-silicon diodes, Solid State Elec. **5**, 285 (1962)

[65] D. J. Chadi, P. H. Citrin, C. H. Park, D. L. Adler, M. A. Marcus and H.-J. Gossman: Fermi-Level-Pinning Defects in Highly n-Doped Silicon, Phys. Rev. Lett. **79**, 4843 (1997)

[66] L. F. Wagner and W. E. Spicer: Photoemission study of the effect of bulk doping and oxygen exposure on silicon surface states, Phys. Rev. B **9**, 1512 (1974)

[67] E. Anderås, L. Vestling, J. Olsson and I. Katardjiev: Resistance electric field dependence and time drift of piezoresistive single crystalline silicon nanofilms, Proc. Chem. **1**, 80 (2009)

[68] W. D. Pilkey: Peterson's stress concentration factors (Wiley-Interscience, New York, 1997)

[69] R. Bashir, A. Gupta, G. W. Neudeck, M. McElfresh and R. Gomez: On the design of piezoresistive silicon cantilevers with stress concentration regions for scanning probe microscopy applications, J. Micromech. Microeng. **10**, 483 (2000)

[70] J. B. Hannon, S. Kodambaka, F. M. Ross and R. M. Tromp: The influence of the surface migration of gold on the growth of silicon nanowires, Nature **440**, 69 (2006)

[71] D. V. Lang, H. G. Grimmeiss, E. Meijer and M. Jaros: Complex nature of gold-related deep levels in silicon, Phys. Rev. B **22**, 3917 (1980)

[72] C. Auth, C. Allen, A. Blattner and D. Bergstrom: A 22nm high performance and low-power CMOS technology featuring fully-depleted tri-gate transistors, self-aligned contacts and high density MIM capacitors, Symp. VLSI Tech. 131 (2012)

[73] E. W. Greeneich and R. S. Muller: Acoustic-wave detection via a piezoelectric field-effect transducer, Appl. Phys. Lett. **20**, 156 (1972)

[74] R. He, X. Feng, M. L. Roukes and P. Yang: Self-transducing silicon nanowire electromechanical systems at room temperature, Nano Lett. **8**, 1756 (2008)

[75] E. Mile, G. Jourdan, I. Bargatin, S. Labarthe, C. Marcoux, P. Andreucci, S. Hentz, C. Kharrat, E. Colinet and L. Duraffourg: In-plane nanoelectromechanical resonators based on silicon nanowire piezoresistive detection, Nanotechnology **21**, 165504 (2010)

[76] P. Singh, J. Miao, V. Pott, W. T. Park and D. L. Kwong: Piezoresistive Sensing Performance of Junctionless Nanowire FET, I.E.E.E. Elec. Dev. Lett. **33**, 1759 (2012)

[77] S. Zhang, L. Lou and C. Lee: Piezoresistive silicon nanowire based nanoelectromechanical system cantilever air flow sensor, Appl. Phys. Lett. **100**, 023111 (2012)

[78] M. Sansa, M. Fernandez-Regulez, A. San Paulo and F. Perez-Murano: Electrical transduction in nanomechanical resonators based on doubly clamped bottom-up silicon nanowires, Appl. Phys. Lett. **101**, 243115 (2012)

[79] P. E. Allain, F. Parrain, A. Bosseboeuf, S. Mâaroufi, P. Coste and A. Walther: Large-Range MEMS Motion Detection With Subangström Noise Level Using an Integrated Piezoresistive Silicon Nanowire, J. Micromech. Sys. **22**, 716 (2013)

[80] T. Iida, T. Itoh, D. Noguchi, and Y. Takano: Residual lattice strain in thin silicon-on-insulator bonded wafers: Thermal behavior and formation mechanisms, J. Appl. Phys. **87**, 675 (2000)

[81] P. E. Allain, X. Le Roux, F. Parrain and A. Bosseboeuf: Large initial compressive stress in top-down fabricated silicon nanowires evidenced by static buckling, J. Micromech. Microeng. **23**, 015014 (2013)

[82] S. W. Chung, J. Y. Yu and J. R. Heath: Silicon nanowire devices, Appl. Phys. Lett. **76**, 2068 (2000)

[83] J. F. Creemer, F. Fruett, G. Meijer and P. J. French: The piezojunction effect in silicon sensors and circuits and its relation to piezoresistance, I.E.E.E. Sensors Journal **1**, 98 (2001)